\documentclass[a4paper,12pt]{article}
\usepackage{graphicx}
\usepackage[dvips]{color}
\usepackage{amsmath}
\definecolor{mycolor}{gray}{.8}
\newcommand{\beeq}{\begin{equation}}
\newcommand{\eneq}{\end{equation}}
\newcommand{\beeqn}{\begin{eqnarray}}
\newcommand{\eneqn}{\end{eqnarray}}

\begin{document}

\vskip 3cm

\begin{center}

{\bf \large{Asymptotic safety of quantum gravity and improved spacetime of black hole singularity by cutoff identification}}\\

\vspace{2.0cm}
\baselineskip=20pt plus 1pt minus 1pt

\vspace{2cm}
{\bf Hiroki Emoto}\footnote[1]{dantes11@msn.com} \\

\vspace{.1cm}
{\it 266-11-303 Nakano-shinden,
Surugaku, Shizuoka 422-8051, Japan}\\

{\it Institute for High Energy Physics, 14220, Protvino, Russia}\\

\end{center}

\begin{abstract}
The possibility of asymptotic safety scenario (asymptotic freedom) for quantum gravity has been pointed out in many contexts recently. From this point of view, we discuss some applications of cutoff identification to the black hole. If we consider the condition that the Newton coupling becomes weaker as approaching the origin, the curvature singularity still remains.   
\end{abstract}

\newpage

\section{Introduction and Motivation}

 General Relativity(GR) has been widely investigated and accepeted as a classical theory of gravity, but some issues are still remained. One of them is the spacetime singularity. This physical singularity happens in quite general situations\cite{Hawking}. Much of the works have been devoted to this problem, but in general, it is expected that the theory of quantum gravity will have to resolve this difficulty in future. Although the theory of quantum gravity has not been established yet, quantum theory may regularize the singularity in some way. Here we discuss about the curvature singularity.   

 Among many candidates for quantum gravity, we concentrate on the "asymptotic safety" scenario which has been investigated for several years recently. This search for a ultraviolet(UV) fixed point of quantum gravity was performed in some methodology and pointed out the possibility of its existence in various dimensions.

 In 4 dimensions, if the UV fixed point actually exists, the coupling of gravity (Newton constant) vanishes in the high energy limit (asymptotic freedom). In the quantum field theory, the high energy limit corresponds to the small distance limit. Therefore, the UV fixed point means that the coupling strength of gravity with matter becomes weaker as we go to shorter distances world, and which will be also equivalent to approaching the center of the gravitational potential field(for example ,in the Schwarzschild spacetime), then the singularity of the curvature of the corresponding spacetime  would become weaker or regularized in some manner intuitively. Especially the information of the UV fixed point (anomalous dimension, fractal dimension) is exactly fixed \cite{Reu-Leu}, which would imply some equivalence of the results from discretized simulation of quantum gravity (dynamical triangulation)\cite{Amb}(Note: The study for the fixed point and phase structure of quantum gravity have also been continuing from the approaches of simplical quantum gravity (see, for instance, \cite{Amb}\cite{Ham-Wil}\cite{Yukawa} and their references)). The weakness of gravity at short distance was also discussed in string theory\cite{Sieg}.  So the connection between the UV limit of quantum gravity and the singularity point of classical spacetime is also an intriguing problem. Even if the gravitational coupling vanishes and no gravity world realize for matter just on the UV fixed point, the asymptotic form of the function of metric or "gravitational coupling" (the degree of  convergence et al) around it are important to judge the existence of singularity.  

In the asymptotic safety scenario, momentum dependence of the Newton constant (or anomalous dimension) is fixed just on the fixed point(if it exists), but we don't know the way of the mapping this information  to the coordinate space in the curved background of geometry. Mainly in this paper, we study along the line taken in works\cite{Reu1}\cite{Reu2}\cite{Reu3}\cite{Reu4}. To study the geometry of the planck scale or in a smaller region, we try to know the form of the metric of the "quantum spacetime" in the coordinate space. 

 In this paper, we discuss prescriptions suggested in some papers \cite{Reu2}\cite{Reu3}\cite{Reu4}, which produced the interesting observational implications for several phenomena. We study the efficiency of  this methods for regularization of the curvature singularity of the black hole. In Section 2, we will review the concept of the nonperturbative renormaliztion group method and its approximated analytical solution from which we will check the effective potential and review the result of \cite{Reu2}. In Section 3, we will practice its improvement which is the same spirit proposed in \cite{Reu3} and discuss whether the central singularity of black hole is regularized or not in these improved spacetime.

\section{Flow solution of Newton coupling}

 It is well known that field theory of gravity cannot be quantized perturbatively because of its nonrenormalizability. According to this difficulty, S.Weinberg suggested another possibility of renormalizability, "asymptotic safety"\cite{Wein}. We introduce its concept below.

 We denote the renormalized coupling parameters as $g_{i}(\mu)$ at a renormalization scale $\mu$. In addition, define the demensionless couplings $\hat{g}_{i}(\mu)$ by $g_{i}(\mu)= \mu^{d_{i}} \hat{g}_{i}(\mu)$. Here $d_{i}$ is the mass dimension of the original couplings.

 Then, any partial or total reaction rate $R$ is written by
\begin{equation}
      R=\mu^{D}f(\frac{E}{\mu}, X, \hat{g}(\mu))\label{reaction}
\end{equation}
Here $D$ is ordinary mass dimension of $R$, and $E$ is some energy characterizing the process, $X$ are all other dimensionless physical variables.
  
Since we can set renormalization scale as $\mu = E$, then (\ref{reaction}) becomes
\begin{equation}
      R=E^{D}f( E, X, g(E))
\end{equation}
 This depends on the behavior of the coupling $g(\mu)$ as $\mu \rightarrow \infty$ apart from the trivial scaling factor $E^D$. Then, we concentrate on the flow equation of the dimensionless coupling of $\hat{g}(\mu)$

The approching of $\hat{g}(\mu)$ to finite value as $\mu\to 0$ means the ultraviolet fixed point of quantum gravity in this case, and he showed it in $(2+\epsilon)$ dimensions by $\epsilon$-expansion. This is essentially the same argument with the modern view point of the Wilson renormalization group approach.

 In higher dimensions, since the Newton constant has a negative mass dimension (UV divergence appears),  we cannot apply usual perturbative computations. Vigorous execution in this direction was started by M.Reuter\cite{Reu1}. Much of the efforts has been dedicated to reveal its aspects from the various view points numericaly and analytically (see, for instance, \cite{Lau-Reu}\cite{Reu-Sau}\cite{Od}\cite{Lit} and references contained therein). In these works, nonperturbative renormalization group approach was applied and a plausible possibility of the existence of non-Gaussian fixed point (UV fixed point) has been pointed out. In this approach, instead of evaluating the each loop contribution,  we evaluate the flow of the effective average action $\Gamma_k$ incorporating the cutoff function $\mathcal{R}_{k}$ which referred the specific cutoff momentum scale $k$, and getting lower the cutoff scale and integrating the lower momentum mode gradually. The flow equation for the efective average action is conceptually
\beeq
k\partial_{k}\Gamma_{k}=\frac{1}{2}\text{Tr}[(\Gamma^{(2)}_{k}+\mathcal{R}_{k})^{-1}k\partial_{k}\mathcal{R}_{k}]
\eneq
 Here $\Gamma^{(2)}_{k}$ is the Hessian of the effective average action $\Gamma_{k}$ itself in field space. Then we derive the flow equation (beta function) of the coupling through the derivative expansion with ansatz of the solution. In \cite{Reu1} and \cite{Reu2}, the Einstein-Hilbert ansatz
\beeq
\Gamma_{k}[g,\bar{g}]=\frac{1}{16\pi G_k}\int d^{d}x\sqrt{g}\{-R(g)+2\bar{\lambda}_k\}+ classical\hspace{0.2cm}gauge\hspace{0.2cm}fixing
\eneq
was studied. Essentially there is no divergence in each step of calculations. $\Gamma_k$ is the standard effective action in the limit $k\to 0$. The stability of the existence of the UV fixed point has been investigated by extending the dimension of the theory space. Here we don't review the details of this formulations any more, and only start from an approximate solution which was already mentioned\cite{Reu2}. In ref\cite{Reu2}, the renormalization group equation was solved without cosomological constant and matter in a cetain approximation

\begin{eqnarray}
G(k)=\frac{G_0}{1+\omega G_0 k^2} ,\label{Newton solution}
\end{eqnarray}
 where $k$ is Euclidean momentum, $\omega$ is a certain constant (order $1$) which is generated from some cutoff functions of momentum integral, $G_0$ is the Newton constant at the reference scale $k_0$ and here we set $k_0\sim 0$(cosmological scale).

 When we take the large momentum limit ($k^2 \gg G_0^{-1}$)
\begin{eqnarray}
G(k)\sim \frac{1}{\omega k^2} ,\label{just on UV}
\end{eqnarray}
 which shows the asymptotic freedom of the gravitational coupling.

 When we consider the phenomenology of this quantum effect, we should translate these information in the momentum space into that for the coordinate space. Exactly speaking, the momentum $k^2$ should be considered as an eingenvalue of the Laplace operator on the general background spacetime. But especially in the curved background case, we don't know the way to pass to the coordinate space via a Fourier integral as we do in the  flat space. Details of discussion about a solution generating approach can be found in\cite{Reu3}\cite{Reu4}.      

\subsection{Correction for gravitational potential in Minkowski space}

 From this subsection we take up the way of "cutoff identification" proposed in \cite{Reu2}, and speculate about that in some example of flat space. 

 Let us concider the behaviour of the static limit of the potential from above solution in coordinate space. As always, we take $k^2\approx |\vec{k}|^2$

\begin{eqnarray}
V(r)=\int\frac{d^3\vec{k}}{(2\pi)^3}e^{-i\vec{k}\vec{x}}\frac{G(k)}{|\vec{k}|^2}=\int\frac{d^3\vec{k}}{(2\pi)^3}\frac{G_0}{|\vec{k}|^2 (1+\omega G_0 |\vec{k}|^2)}e^{-i\vec{k}\vec{x}}\nonumber\\
=\int\frac{d^3\vec{k}}{(2\pi)^3}G_0\left(\frac{1}{|\vec{k}|^2} -\frac{1}{|\vec{k}|^2 + m_{pl}^2}\right)e^{-i\vec{k}\vec{x}}=\frac{G_0}{4\pi r}(1-e^{-\frac{m_{pl}}{\sqrt{\omega}}r})\label{Newton corrected potential}
\end{eqnarray}

 Here $m_{pl}^2\equiv \frac{1}{G_0}$. This correciton term is of Yukawa type, and it was also found in another method "YFS resummation"\cite{Ward}.In addition, the effective propagator of our case is the inverse of 4th order of momentum which corresponds to $R^2$ (quadratic curvature) theories of perturbative quantum gravity. We note here interesting remark that elaborated work of $R^2$ - background metric independent formalism\cite{Hamada} is related to the results of computer simulation of simplical quantum gravity\cite{Yukawa} (The second term in the third equation (\ref{Newton corrected potential}) is of the same form as massive propagator of negative sign, which has been also known in quadratic curvature's quantum gravity. About this problem of quadratic curvature and conformal factor's divergence in Euclidean gravity in the original formulation of this nonperturbative approach, we can refer to \cite{Lau-Reu} as one observation for the convergence of effective action. We don't proceed in this issue here). From this result, we define the coordinate-dependent Newton constnant

\begin{eqnarray}
G(r)\equiv G_0(1-e^{-\frac{m_{pl}}{\sqrt{\omega}}r})\label{coord Newton}
\end{eqnarray}   

 Certainly $G(r)$ vanishes as $r\rightarrow 0$. This implies asymptotic freedom.

 Above consideration were all in flat space. If one wants to know the quantum correction of the classical solution (we discuss mainly about the Schawarzschild spacetime in this paper) by making use of the momentum flow solution of the Newton constant(\ref{Newton solution}), we have to consider and integrate them in covariant manner with respect to the background spacetime. Actually in the original method of this approach, we use the formula of heat kernel to evaluate flow of the effective action on the maximally symetric space\cite{Reu1}. However it is still difficult to evaluate for general curved background case including the flow solution of the coupling.

 One alternative prescription was proposed and investigated in\cite{Reu2}. In \cite{Reu2}, speculating from the example of QED quantum correction for electric charge, authors assumed some identification rule between momentum cutoff and corresponding cutoff in coordinate space. It is the direct substitution
\begin{equation}
k\Rightarrow\frac{\xi}{r}
\end{equation}
into the coupling flow solution for the momentum space. Here $\xi$ is some constant which will be determined later and $r$ is the distance from the origin. Generally such relation for identification does not hold. We comment a little about it in the appendix. But this is the case for usual transoformation into coordinate representation from momentum space (momentum Fourier integration). 

  From the original spirit of the effective average action, they tried to mimic the  behaviour for inclusion of nonlocal terms \cite{Reu3}\cite{Reu4}. Actually effective action can have nonlocal terms, and the case for momentum cutoff theory was mentioned. In this case, the effective average action generates nonlocal terms during the evolution even if original one consists of local terms at UV scale. As a result, this identification is expected to be the correct behaviour even at the tree level  As application for the infrared physics, dark matter problem of galactic rotation curve was discussed by assuming the flow solution for infrared fixed point of quantum gravity\cite{Reu4}. Cosmological solution was also discussed in \cite{Reu3}.  

 So we follow this identification program in this paper and study the solution around the origin of the "black hole" solution.

\subsection{Cutoff identification in black hole background}

 We review the application for the black hole \cite{Reu2}. This prescription of identification was revised in recent works\cite{Reu3}\cite{Reu4}, which we will discuss about later. To try to generalize above operation, the choices of the invariant length $d(r)$ was adopted instead of the distance $r$ itself which has no universal meaning in curved background.          
 
 In the Schwarzschild case, the authors of \cite{Reu2} indentified
\begin{eqnarray}
k(P)=\frac{\xi}{d(P)}   , \label{cutoff}
\end{eqnarray} 
 here $P$ is the point of reference scale and $d(P)$ is the proper length  between the center of the black hole and point $P$. $\xi$ is some numerical constant to be fixed later.Having integrated the line element with fixed angular and time coordinate, we get the concrete expression of $d(P)$, which is $d(r) \approx r$ as $r\rightarrow \infty$ and $d(r)\propto r^{\frac{3}{2}}$ as $r\to 0$. This analytical expression are different each other whether the refrence point $P$ is inside or outside the horizon.
 
  As a candidate of cuttoff function for interpolating this behaviour in each limit,  
\begin{eqnarray}
d(r)=\left(\frac{r^3}{r+\gamma G_0 M}\right)^{\frac{1}{2}}
\end{eqnarray} 
 was used. Here $\gamma = \frac{9}{2}$ but this was also used as free parameter.Then, we get by replacement (\ref{cutoff})
\begin{eqnarray}
G(r)=\frac{G_0 d(r)^2}{d(r)^2 + \tilde{\omega}G_0}=\frac{G_0 r^3}{r^3 + \tilde{\omega G_0 (r+\gamma G_0 M)}}
\end{eqnarray}
 Here $\tilde{\omega}\equiv \omega\xi^2$. For large distances,the leading correction to Newton constant is given by
\begin{equation}
G(r)=G_0 - \tilde{\omega}\frac{G_0^2}{r^2}+\mathcal{O}\left(\frac{1}{r^3}\right)\label{Newton correction 1}
\end{equation}
and at small distances 
\begin{equation}
G(r)=\frac{r^3}{\gamma \tilde{\omega}G_0 M} + \mathcal{O}(r^4)\label{Newton correction 2}
\end{equation}
 Newton coupling vanishes at $r\to 0$ in this case also. Then we can determine the value $\xi$ at this stage. From the view point of the low energy effective field theory, we can evaluate the quantum corrections to the Newton potential from the perturbative quantum gravity at large distances\cite{Don}. The dominant terms in this case generate as non-analytic terms of momentum. Integrating Fourier integrals, the dominant correction of order $\frac{1}{r^3}$ appears. So  our identification procedure leads to the same order of dominant correction with \cite{Don} actually.  By adjusting the numerical value of (\ref{Newton correction 1}) to the result of \cite{Don} 
\begin{equation}
\tilde{\omega} \equiv \frac{118}{15\pi}
\end{equation}        

 The sructure of this quantum black hole was investigated closely by assuming the lapse function as 
\begin{equation}
f(r)=1-\frac{2G(r)M}{r}\label{lapse}
\end{equation}  
As we see clearly from the expression (\ref{Newton correction 2}), the value of the metric coefficient $f$ is zero at $r=0$ and actually this spacetime has two horizon. When the mass of black hole is lower than a certain characteristic scale (which is comparable to planck mass), horizon disappears and there is no evaporation of the Hawking process.

 Here we comment about the curvature behaviour at origin of this geometry. The metric 
\begin{equation}
ds^2 = -f(r)dt^2 + f(r)^{-1}dr^2 + r^2 d\Omega^2
\end{equation}
with lapse ($c, \nu$ are constants)
\begin{equation}
f(r)=1-c r^{\nu}
\end{equation}
has the curvature and Weyl invariants
\begin{eqnarray}
R=c(\nu +1)(\nu +2)r^{\nu -2}\label{Ricci scalar}\\
R_{\mu\nu\rho\sigma}R^{\mu\nu\rho\sigma}=c^2 (\nu^4 -2\nu^3 +5\nu^2 +4) r^{2\nu -4}\label{Riemann}\\
C_{\mu\nu\rho\sigma}C^{\mu\nu\rho\sigma}=\frac{c^2}{3}(\nu -1)^2(\nu -2)^2 r^{2\nu -4}\label{Weyl}
\end{eqnarray}
 The classical Schwarzschild metric is $\nu=-2$ and (\ref{Riemann}),(\ref{Weyl}) diverges at $r=0$ although Ricci scalar (\ref{Ricci scalar}) is zero. 
 On the other hand, the corrected spacetime has asymptotic form $f=1-2(\gamma\tilde{\omega}G_0)^{-1}r^2 + \mathcal{O}(r^3)$, whose dominant term resembles the de-Sitter metric. Therefore this case with exponent $\nu=2$ has no curvature singularity at $r=0$. 

 Although these results are intriguing, the above prescription should be revised because we should measure the length in the "corrected" space and not in the classical one. In addition, there may be another problem. If the corrected solution of the gravitational system is largely deviated from the classical solution, it may imply that we should totaly reconsider from the very beginning of (classical) action principle. We discuss the revised approach for these two problems in the next section.

\section{"Consistent cutoff identification approaches" to the black hole spacetime}

\subsection{Solution improvement by consistent cutoff identification}

 As we have reviewed already,in ref \cite{Reu3}, we measure the proper distance from the center of the black hole to the "observer" through the metric of the classical solution. But If we obey to this spirit faithfully, proper length should be measured by the metric of quantum corrected spacetime itself. This corresponds to including the backreaction of quantum gravity effects. This point was also mentioned as "consistent cutoff identification" already \cite{Reu4}, but it was not applied rather to the cosomological background or infrared scale than to the black hole or short distance scale.

 So let's perform this program in our black hole case.      
We start from the lapse function (\ref{lapse}) to measure the proper distance.Then the distance function is 
\begin{eqnarray}
 d(r)=\int \sqrt{|ds^2|} = \int^r_0 dr'\sqrt{\Big|\frac{r'}{r'-2G(r') M}\Big|}\label{new distance}
\end{eqnarray}
 Here we replace the Newton constant $G_0$ of \cite{Reu2} with some r-dependent function, $G(r)$. On the other hand, the flow of the Newton coupling in the coordinate expression is determined by
\begin{equation}
G(r)=\frac{G_0 d(r)^2}{d(r)^2 + \omega\xi^2 G_0} \label{flow coord}
\end{equation}  
 To solve both equations (\ref{new distance}) and (\ref{flow coord}) simaltaneously, we differentiate each equations with $r$, and equating 
\begin{eqnarray}
\frac{1}{2}\sqrt{\frac{G_0-G(r)}{\tilde{\omega}G_0 G(r)}}\frac{\tilde{\omega}G_0^2 G^{'}(r)}{(G_0-G(r))^2} = \sqrt{\Big|\frac{r}{r-2G(r) M}\Big|}
\end{eqnarray}
and rewriting this as
\begin{eqnarray}
\frac{1}{4}\tilde{\omega}G_0^3 \left(\frac{dG}{dr}\right)^2 = \Big|\frac{r}{r-2G(r)M}\Big|G(r)(G_0-G(r))^3 ,\label{consistent eq}
\end{eqnarray}
we seek the solution near the origin $r=0$. We assume it to scale as $G(r)\sim r^{\nu}$ $(\nu >0)$ in the limit $r\to 0$. The assumption $\nu>0$ means that $G(r)\to 0$ as $r\to 0$. This is not derived exactly from theory but follows from the intuitive picture of "asymptotic safety (asymptotic freedom)" and the assumption $k=\frac{\xi}{d(r)}$ of (\ref{cutoff}), that is, the energy scale becomes higher as $r\to 0$.

 If we suppose $0<\nu<1$ then the left hand side of (\ref{consistent eq}) is divergent as $r\to 0$ but the right hand side vanishes. So this is not a solution. Now, for the case $\nu\geq 1$ the dominant order is found to be $r^2$ and 
\begin{equation}
G(r)=\frac{1}{\tilde{\omega}}r^2 + \frac{M}{\tilde{\omega}^2}r^3 + \mathcal{O}(r^3)
\end{equation}       
 which leads to the following lapse function
\begin{equation}
f(r)=1-\frac{2M}{\tilde{\omega}}r -2\left(\frac{M}{\tilde{\omega}}\right)^{2}r^2+\dots
\end{equation}
 this corresponds to $\nu=1$ in (\ref{Ricci scalar})(\ref{Riemann}) and these curvature invariant diverge as $r\to 0$. The curvature singularity still remains intact.

\subsection{Improved action principle and Weyl symmetry}

 In this subsection, we proceed the discussion along the solution generating technic proposed in \cite{Reu3}. This method is based on the interesting observation of the Weyl symmetry derived from the new framework of a position-dependent Newton coupling action. As we mentioned before, if the quantum corrected solution is modified largely from the classical one (this is the case near the origin, the previous section), this effect can modify the stationary point of the classical action again. This means a certain kind of back reaction of quantum effects at a much more profound level. We try to solve the problem self-consistently at the action level.We review this method in short.

  The running Newton coupling means that the Newton coupling is position dependent. If we take it seriously in classical level, we have to change the procedure of the use of the action principle for the original Einstein-Hilbert action. This viewpoint is similar the well known Brans-Dicke theory (scalar-tensor gravity) in which the Newton constant is considered to be a position dependent scalar function. In our case also, the equation of motion is changed and to satisfy the Bianchi identity, we add a new term $S_{\theta}[g_{\mu\nu},G,\Lambda]$ from the beginning. So in the following formulation, the Newton coupling is not independent degree of freedom but is controlled by the constraint and the renormalization group.

 We replace the Newton and cosmological constants with functions $G(x)$ and $\Lambda(x)$ in the Einstein-Hilbert action 
\begin{eqnarray}
S_{mEH}[g_{\mu\nu},G,\Lambda]=\frac{1}{16\pi}\int d^4x\sqrt{-g}\frac{1}{G(x)}\left(R-2\Lambda(x) \right)
\end{eqnarray}
The equation of motion from this action is
\begin{equation}
\frac{2}{\sqrt{-g}}\frac{\delta S_{mEH}}{\delta g^{\mu\nu}}=\frac{1}{8\pi G}(G_{\mu\nu}+\Lambda g_{\mu\nu}-\triangle t_{\mu\nu}) ;
\end{equation}
here $\triangle t_{\mu\nu}$ is the outcome from position dependent Newton coupling $G(x)$
\begin{eqnarray}
\triangle t_{\mu\nu}=G(x)(\nabla_{\mu}\nabla_{\nu} - g_{\mu\nu}\nabla^2)\frac{1}{G(x)}\hspace{5cm}\nonumber\\
=\frac{1}{G^2}\left(2\nabla_\mu G \nabla_\nu G -G\nabla_\mu \nabla_\nu G - g_{\mu\nu}(2g^{\rho\sigma}\nabla_\rho G \nabla_\sigma G -G\nabla^2 G)\right)
\end{eqnarray}
 covariant derivative $\nabla_\mu$ is taken with respect to the metric $g_{\mu\nu}$. Then we consider the energy momentum tensor $T^{\mu\nu}$ of matter with the conservation law $\nabla^{\mu}T_{\mu\nu}=0$. However, the Bianchi condition $\nabla^{\mu}G_{\mu\nu}=0$ requires this equation of motion to hold identically (irrespicitvely of its solution), and we should add new compenesating term $\theta_{\mu\nu}$. Finally the total equation of motion is 
\begin{eqnarray}
G_{\mu\nu}= -\Lambda g_{\mu\nu}+\triangle t_{\mu\nu} +8\pi G(T_{\mu\nu}+\theta_{\mu\nu}) \label{eq of new action}
\end{eqnarray} 
Contracting this with the covariant derivative, and using $\nabla^{\mu}G_{\mu\nu}=0$ and $\nabla^{\mu}T_{\mu\nu}=0$, we get the "on-shell consistency condition"
\begin{eqnarray}
\frac{3}{2G^3}{G\nabla^2 G -2\nabla_\rho G \nabla^{\rho}G} + \nabla^{\mu}\vartheta_{\mu\nu} -\frac{1}{G}(\nabla^{\mu}G) \tilde{\vartheta}_{\mu\nu}\nonumber\\
+ 4\pi T(\nabla_\nu G)-\frac{1}{G}\nabla_\nu (G\Lambda)=0 \label{consistency condition}
\end{eqnarray}   
with definitions $\vartheta_{\mu\nu}\equiv 8\pi G\theta_{\mu\nu}$ and $\tilde{\vartheta}_{\mu\nu}\equiv \theta_{\mu\nu} - \frac{1}{2}g_{\mu\nu} \theta^{\rho}_{\rho}$, $T \equiv T^{\rho}_{\rho}$. 

 We can not solve (\ref{consistency condition}) for $\vartheta_{\mu\nu}$ in general case. But for a special case,$T=0$, at least, when matter $A$ satisfy its equation of motion, and $G(x)\Lambda(x)=constant$, we solve the equation under the assuumption that $\vartheta_{\mu\nu}$ vanishes for $\psi=constant$
\begin{eqnarray}
\vartheta_{\mu\nu}=-\frac{3}{2G^2}[\nabla_\mu G \nabla_\nu G -\frac{1}{2}g_{\mu\nu}\nabla_{\rho}G \nabla^{\rho}G ]
\end{eqnarray}
 here $\psi\equiv -\ln(\frac{G(x)}{G_0})$. This tensor $\theta_{\mu\nu}$ is derived from 
\begin{eqnarray}
S_\theta =\frac{3}{32\pi G_0}\int d^4 x \sqrt{-g}e^{\psi}\nabla_\mu \psi \nabla^{\mu}\psi
\end{eqnarray}
 by variation, $\theta_{\mu\nu}=\frac{2}{\sqrt{-g}}\frac{\delta S_\theta}{\delta g_{\mu\nu}}$. 
Total action with matter, in the end, is
\begin{eqnarray}
S_{tot}=S_{mEH}[g_{\mu\nu},G,\Lambda]+S_M[g_{\mu\nu},A]+S_{\theta}[g_{\mu\nu},G,\Lambda]
\end{eqnarray}
here  $A$ is  matter fields in the matter action $S_M$. The equation of motion obtained by variation of the metric is (\ref{eq of new action}).

 From here, we consider the case without position-dependent cosmological constant, $\Lambda=0$, and without matter, $T_{\mu\nu}=0$.
\begin{equation}
S_{mEH}[g,G,0] +S_\theta^{BD}[g,G]=\frac{1}{16\pi G_0}\int d^4 x\sqrt{-g} e^{\psi}[R(g)+\frac{3}{2}\nabla_\mu \psi \nabla^\mu \psi]
\end{equation}
 This action can be transformed into the usual Einstein-Hilbert action of a new metric $\gamma_{\mu\nu}$ by conformal transformation $\gamma_{\mu\nu}\equiv e^\psi g_{\mu\nu}=\frac{G_0}{G(x)}g_{\mu\nu}$ 
\begin{equation}
S_{mEH}[g,G,0] +S_\theta^{BD}[g,G]=S_{EH}[\gamma]=\frac{1}{16\pi G_0}\int d^4 x\sqrt{-g}R(\gamma) , 
\end{equation}
 up to a total derivative term.

The existence of this symmetry shows that the stationary point (classical solution) of the original action wiht the metric $g_{\mu\nu}$ corresponds to that of the Einstein-Hilbert action with the metric $\gamma_{\mu\nu}$. We can make use of this symmetry to solve the problem.

 We search for a static vacuum solution with spherical symmetry. From here we relabel the metric of the Einstein-Hilbert system as $\gamma_{\mu\nu}\to \bar{g}_{\mu\nu}$, and call each system $g_{\mu\nu}$, $\bar{g}_{\mu\nu}$ as "Q-system" and "C-system" respectively. They are conformally related $g_{\mu\nu}=(\frac{G(x)}{G_0})\bar{g}_{\mu\nu}$. In the C-system, the static spherically symmetric solution is the Schwarzschild solution, therefore the metric in the Q-system is  
\begin{eqnarray}
ds^2=\frac{G(r)}{G_0}d\bar{s}^2=\frac{G(r)}{G_0}\left(-f(r)dt^2+f(r)^{-1}dr^2+r^2 d\Omega^2 \right) \label{conformal metric}
\end{eqnarray}   
here
\begin{equation}
f(r)=1-\frac{2G_0 M}{r}
\end{equation}
 Concrete form of the function $G(r)$ follows from the same identification procedure as in the previous subsection. The distance function from this metric is
\begin{eqnarray}
 d(r)=\int \sqrt{|ds^2|} = \int^r_0 dr' \sqrt{\left(\frac{G(r')}{G_0}\right)\Big|\frac{r'}{r'-2G_0 M}\Big|} ; \label{conformal distance}
\end{eqnarray} 
 the flow equation is the same as (\ref{flow coord}). So the master equation corresponding to (\ref{consistent eq}) is
\begin{eqnarray}
\frac{1}{4}\tilde{\omega}G_0^4 \left(\frac{dG}{dr}\right)^2 = \Big|\frac{r}{r-2G_0 M}\Big|G(r)^2(G_0-G(r))^3 \label{conformal consistent eq}
\end{eqnarray}
 As is the case in the previous section, we assume the scaling $G(r)\sim r^\nu$ and search for an analytic solution of this equation around $r=0$. When we assume $0<\nu<1$, the left hand side of (\ref{consistent eq}) is divergent as $r\to 0$ but the right hand side vanishes. Again this is not solution. Then we study the case $\nu \geq 1$. However in this case, the lowest order of right hand side is $r^{2\nu +1}$ although the left hand side is $r^{2\nu-2}$. This shows that there is no analytic solution around the origin $r=0$.

To demonstrate it more clearly, we consider (\ref{conformal consistent eq}) again in the limit $r\to 0$. Eq.(\ref{conformal consistent eq}) reduces approximately to  
\begin{equation}
\frac{\sqrt{\tilde{\omega}}}{2}{G_0}^2\frac{dG}{dr}=\sqrt{r}G(r)(G_0 - G(r))
\end{equation}
Integrating this in the region $0\leq r'\leq r$ we get
\begin{equation}
\frac{\sqrt{\tilde{\omega}}}{2}{G_0}\int^G_0 dG' \left(\frac{1}{G_0 -G'}+\frac{1}{G'} \right)=\int^r_0 dr' \sqrt{r'}
\end{equation}
while the integration of the left hand side of the above equation is logarithmic divergent at  $G'=0$ . The solution can not be expressed with any power function or analytic solution of $r$ around the origin.

 By the way, we can transform the metric (\ref{conformal metric}) into the standard radial coordinate form by $\rho(r) \equiv r\sqrt{\frac{G(r)}{G_0}}$
\begin{equation}
ds^2= -\frac{G(r)}{G_0}f(r)dt^2 +\frac{d\rho^2}{f(r)(1+\frac{r}{2}\ln \frac{G}{G_0})}+\rho^2 d\Omega^2 ,
\end{equation}
except we interpret $r=r(\rho)$. If we look for an analytic solution $G(\rho)$ with same prescripton as above  around the origin $\rho=0$, assuming $G(\rho)=\rho^\nu (\nu > 0)$ at $\rho\to0$, we can't find a suitable solution for $\nu$. There is no desired solution in this coordinate frame also.

 To see the aspect of curvature singularity, we go back to a more general point of view. Curvature invariants transform according to the conformal transformation $g_{\mu\nu} \to \bar{g}_{\mu\nu}=\Omega^2(x)g_{\mu\nu}$ 
\begin{eqnarray}
\bar{R}=\Omega^{-2}R - 6 \Omega^{-3}\nabla^2 \Omega \hspace{7.5cm}\\
\bar{R}_{\mu\nu\rho\sigma}\bar{R}^{\mu\nu\rho\sigma}=\Omega^{-4}R_{\mu\nu\rho\sigma}R^{\mu\nu\rho\sigma}-8\Omega^{-5}R^{\mu\nu}\nabla_{\mu}\nabla_{\nu}\Omega\hspace{3cm}\nonumber\\ 
+\Omega^{-6}(16R^{\mu\nu}\nabla_{\mu}\Omega\nabla_{\nu}\Omega -4R\nabla_{\mu}\Omega\nabla_{\mu}\Omega +16\nabla_{\mu}\Omega\nabla_{\nu}\Omega\nabla~{\mu}\nabla^{\nu}\Omega \nonumber\\
-8\nabla_{\mu}\Omega\nabla_{\nu}\Omega\nabla^{\nu}\nabla^{\mu}\Omega+4\nabla^2\Omega\nabla^2\Omega ) \nonumber\\
+\Omega^{-7}(-32\nabla_{\mu}\Omega\nabla_{\nu}\Omega\nabla^{\mu}\nabla^{\nu}\Omega + 8\nabla_{\mu}\Omega\nabla^{\mu}\Omega\nabla^{2}\Omega)
 -8\Omega^{-8}(\nabla_{\mu}\Omega\nabla^{\mu}\Omega)^2
\end{eqnarray}
 On the other hand, the Weyl tensor is invariant under general conformal transformations.

 From these expression, if we assume $G(r)\to 0$ at $r\to 0$ as in the previous section, the curvature still diverges at the origin.
 
\section{Discussion and Conclusion}

 We have tried to get some sort of description for quantum corrected geometry from the view point of asymptotic safety. We started from some approximate solution for the running equation.

Most simple description is that in Mikowski spacetime. 
In this case, we can get an analytical solution from the approximate solution simply by the Fourier "transformation". This solution shows that the effective Newton coupling dissapears when we approach the origin. If we adhere to this picture, situation is simply asymptotic freedom in the high energy (short distance) limit. However, this rather corresponds to the quantum correction for the Newton potential.

If we examine the quantum correction for a realistic configuration which is described by general relativity, we have to find the reasonable scheme for "transformation" which produce the consistent results with observational data or other theoretically well-defined results.
The result from the first proposal in \cite{Reu2} conformed to that of the effective field calculus surely at long distances. We have discussed revised way of a cutoff identifications.  

 Our interest has been focused on the singular spacetime (curvature singularity), and we adopted two ways of consistent cutoff identification to the static spherically symmetric spacetime. We have studied the solution $G(r)$ under condition of intuitive asymptotic freedom, $G(r)\to 0$ at $r\to 0$ as an analytical solution.  

 First approach is the solution improvement. This is the proposal which we fix the solution form for the Newton coupling and we measure the reference length on this solution. In this case, we got an analytical solution for $G(r)$, but the curvature singularity still remains although its strength became weaker than the classical one.

 Second approach, the action improvement, is based on a serious attention to the meaning of the running coupling. In this framework we can utilize the special feature of the "new action", conformal symmetry, for some kind of spacetime. That is our case. In this case we could not integrate the solution and it seems to be non-analytic. Curvature singularity also persists. For this second approach, we should remark that  even if we leave the "cutoff identification" regime, the curvature divergence is not regularized with condition $G(r)\to 0$ at $r\to 0$.

 Consequently our approach by the cutoff identification which was mainly proposed in \cite{Reu2}\cite{Reu3} does not regularize the curvature singularity at the center of the black hole. The "cutoff identification" is phenomenological approach at present in this sense. (This situation seems to be the same in cosmology, where there is no universal observation to regularize the initial time singularity 'Big Bang' in much works (see \cite{Reu3}\cite{Reu5} and so on)).If we give up the condition $G(r)\to 0$ at $r\to 0$, the situation would change, but in this case we should proceed much more carefully.

 On the other hand, the second approach, i.e. the action improvement or Weyl symmetry may give us some interesting insights. If we accept this way, the singularity still appears under the condition $G(r)\to 0$ at $r\to 0$ without particular cutoff identification. 

 As we will see in the appendix, the speculation of just on the UV fixed point (ex.(\ref{Newton solution})) and that of including the next order correction (ex.(\ref{just on UV})) lead to different asymptotic behaviour in the limit $r\to 0$ in the usual integral of flat space. The results in the appendix also imply that naive manipulations for approximation does not surely produce correct result.
To describe the situations near the UV fixed point as some geometry seems to be difficult at present. The construction of "quantum geometry" around the UV fixed point from the "renormalization group" at short distance scale still needs to be studied carefully. 

 We also note another works for running Newton coupling of quantum gravity effect which was applied to cosmology\cite{Ham},  but this method is also suitable for infrared scale physics from the beginning. The connections of "asymptotic safety" and string theory was also discussed recently in the search for bouncing universe\cite{Sieg2}.  

 In this paper, we studied the expectations of the regularization of singularity for black hole by quantum gravity effects. There are still other perspectives for this problem. For instance, Weyl curvature hypothesis\cite{Pen} by R.Penrose insists the distinguishment of the singularity in Black hole with that of Big Bang cosmology. Generation of the time-irreversibility may be also interesting problem for quantum gravity from that point of view.  

{\Large{\bf{Acknowledgement}}}

 I would like to thank A.A.Logunov, V.A.Petrov, A.P.Samokhin for hearty invitation and helpful discussion in the theoretical division of IHEP(Protvino). This work was based on the scientific grant from JINR(Dubuna). In addition, I am grateful to T.Kubota for his encouraging discussion in the early stage of this study at Osaka University.

{\Large{\bf{Appendix}}}

In this appendix, we see the situation of Fourier integral of potential and cutof identification in the renormalization flow of coupling, and how it works. This prescription was used from the slightly differnt point of view in recent works\cite{Reu3}\cite{Reu4}, which I have already explained in this paper. 
 
 The prescription proposed and tested in \cite{Reu2}, below we review. As is well known, the one loop corrected potential (Uehling potential) in QED has the form 
\begin{eqnarray}
V(r)=-\frac{e^2}{4\pi r}\left(1+\frac{e^2}{6\pi}\int_{2m}^{\infty}dq\frac{e^{-qr}}{q} \sqrt{1-\frac{4m^2}{q^2}}\left(1+\frac{2m^2}{q^2} \right)\right) 
\end{eqnarray}
We can estimate this potential in two asymptotic regions:
\begin{eqnarray}
V(r)=-\Large{\frac{e^2}{4\pi r}}\left(1+\frac{e^2}{4 \pi^{3/2}}\frac{1}{(mr)^{3/2}}e^{-2mr} +\dots \right) \hspace{1cm} (mr \gg 1)
\end{eqnarray}
\begin{eqnarray}
V(r)=-\frac{e^2}{4\pi r}\left(1+\frac{2 e^2}{3 \pi}\ln(\frac{1}{mr})-\dots \right) \hspace{3cm} (mr \ll 1 ) \label{approx}
\end{eqnarray}
On the other hand, the running gauge coupling is
\begin{equation} 
{\Large{e^2(k)}}=\frac{e^2}{1-\frac{e^2}{6\pi^2}\ln{\frac{k}{k_0}}}\label{electric}
\end{equation}
 The replacement $k\leftrightarrow\frac{1}{r}$ in (\ref{electric}) restores the correction term of the potential(\ref{approx}) in the case $\frac{e^2}{6\pi^2}\ln{\frac{k}{k_0}} \ll 1$, up to a constant coefficient.

 So this operation is considered here as the approximate estimation of the functional form of the dominant contribution from quantum corrections. From this observation, authors tried to access to the solution of quantum corrected spacetime by generalizing above replacement to the curved space ingredients $k\leftrightarrow \frac{1}{d(r)}$, here $d(r)$ is the invariant length in curved space for a specific reference scale. Before explaining this cutoff identification, we see the second example of this replacement 

 Let's see the case of the gravitational correction again to check how this operation can be worked out. If we start from the flow solution of the Newton constant(\ref{Newton solution}), the result of coordinate representation(\ref{Newton corrected potential}) and (\ref{coord Newton}) is exact.

 In the small distance limit $r\ll m_{pl}^{-1}$ 
\begin{eqnarray}
G(r)\approx G_0\left(\frac{m_{pl}}{\sqrt{\omega}}r+\frac{1}{2}\left(\frac{m_{pl}}{\sqrt{\omega}}\right)^2 r^2 + \mathcal{O}(r^3) \right) 
\end{eqnarray}   
 The leading term is linear in distance.

 On the other hand, according to the identification $k\leftrightarrow \frac{1}{r}$ in the solution in the momentum space (\ref{Newton solution}), rewriting the $G(k)$ with $G(r)$
\begin{eqnarray}
G(r)=\frac{G_0 r^2}{r^2 + \omega G_0^2}\approx \frac{r^2}{\omega}-\frac{r^4}{\omega^2 G_0} + \frac{r^6}{\omega^3 G_0^2} - \dots \label{replace sol}
\end{eqnarray}   
 in the small distance limit. Here the leading term is quadratic. That is, the replacement does not recover the right behaviour of the funciton in the leadin order. The reason for which we pay a careful attention to the leading order of the distance in small distance limit is that it relates the strength of curvature divergence as we will see in the eq.(\ref{Ricci scalar}) et al. (To see the degree of this central divergence, coefficient of the leading order is not essential) 

When we expand the solution (\ref{Newton solution}) in the limit $|\vec{k}|^2 \gg m_{pl}^{-2}$ from the biginning
\begin{eqnarray}
V(r)=\frac{1}{\omega}\int\frac{d^3\vec{k}}{(2\pi)^3}\frac{e^{-i\vec{k}\vec{x}}}{|\vec{k}|^4} \left(1 - \frac{1}{\omega G_0 |\vec{k}|^2} + \frac{1}{(\omega G_0)^2}\frac{1}{|\vec{k}|^4} - \dots \right)\nonumber\\
= \frac{1}{4\pi\omega}\left(\frac{r}{2} - \frac{1}{\omega G_0}\frac{r^3}{4!} + \frac{1}{(\omega G_0)^2}\frac{r^5}{6!}- \dots \right) \label{approx integration}
\end{eqnarray} 
 the effective coupling is
\begin{eqnarray}
G(r)\approx \frac{r^2}{2\omega}-\frac{r^4}{4! \omega^2 G_0} + \frac{r^6}{6! \omega^3 G_0^2} - \dots
\end{eqnarray}
 The powers of this behaviour coincide with that of (\ref{replace sol}) up to constants. The reason of this coincidence is that the result of the momentum integral of the power function in $|\vec{k}|$ with Fourier factor is also simple function of the distance after the suitable regularization, and its behaviour can be predicted only from the dimensional consideration up to numerical factors. For instance, $\int\frac{d^3\vec{k}}{(2\pi)^3}\frac{1}{|\vec{k}|^4}e^{-i\vec{k}\vec{x}} \sim r$. So if the integrand of the above Fourier integral is written in terms of a polynominal series in $|\vec{k}|$, its leading behaviour is coincident with the replacement solution by $k\leftrightarrow \frac{1}{r}$ at least up to a constant coefficient. On the other hand , when its integrand is not a polynominal series, for example (\ref{Newton solution}), precise evaluation does not coincide with the outcome from the simple replacement $k\leftrightarrow \frac{1}{r}$ in the leading order of the small distance behaviour.

-------------------------------------------------------------------------------

\def\jnl#1#2#3#4{{#1}{\bf #2} (#4) #3}

\def\Zphys{{\em Z.\ Phys.} }
\def\jssc{{\em J.\ Solid State Chem.\ }}
\def\jpsJ{{\em J.\ Phys.\ Soc.\ Japan }}
\def\ptps{{\em Prog.\ Theoret.\ Phys.\ Suppl.\ }}
\def\PTP{{\em Prog.\ Theoret.\ Phys.\  }}

\def\JMP{{\em J. Math.\ Phys.} }
\def\NPB{{\em Nucl.\ Phys.} B}
\def\NP{{\em Nucl.\ Phys.} }
\def\PLB{{\em Phys.\ Lett.} B}
\def\PL{{\em Phys.\ Lett.} }
\def\PRL{\em Phys.\ Rev.\ Lett. }
\def\PRB{{\em Phys.\ Rev.} B}
\def\PRD{{\em Phys.\ Rev.} D}
\def\PRe{{\em Phys.\ Rep.} }
\def\AP{{\em Ann.\ Phys.\ (N.Y.)} }
\def\RMP{{\em Rev.\ Mod.\ Phys.} }
\def\ZPC{{\em Z.\ Phys.} C}
\def\SCI{\em Science}
\def\CMP{\em Comm.\ Math.\ Phys. }
\def\MPLA{{\em Mod.\ Phys.\ Lett.} A}
\def\IJMPB{{\em Int.\ J.\ Mod.\ Phys.} B}
\def\IJMPA{{\em Int.\ J.\ Mod.\ Phys.} A}
\def\PR{{\em Phys.\ Rev.} }
\def\cmp{{\em Com.\ Math.\ Phys.}}
\def\JPA{{\em J.\  Phys.} A}
\def\CQG{\em Class.\ Quant.\ Grav. }
\def\ATMP{{\em Adv.\ Theoret.\ Math.\ Phys.} }
\def\ibid{{\em ibid.} }
\def\Poi{{\em Ann.\ Inst.\ H.\ Poincare}}

\leftline{\Large{\bf References}}

\renewenvironment{thebibliography}[1]
        {\begin{list}{[$\,$\arabic{enumi}$\,$]}  
        {\usecounter{enumi}\setlength{\parsep}{0pt}
         \setlength{\itemsep}{0pt}  \renewcommand{\baselinestretch}{1.2}
         \settowidth
        {\labelwidth}{#1 ~ ~}\sloppy}}{\end{list}}

\end{document}